\documentclass{agnspec}
\usepackage{graphics}
\usepackage{epsfig}

\begin{document}
\title{How Massive are BALQSO Winds?}
\author{F. Hamann\inst{1}, B. Sabra\inst{1} \and
V. Junkkarinen\inst{2}, R. Cohen\inst{2} \and G. Shields\inst{3}}
\institute{University of Florida, USA \and
University of California, San Diego, USA \and University of Texas,
Austin, USA}
\maketitle

\begin{abstract}

We are involved in a program to derive properties of broad absorption
line (BAL) winds in quasars using combined UV and X-ray observations.
A major obstacle is large uncertainties in the derived BAL column
densities because of partial coverage of the background light source.
In this preliminary report, we circumnavigate those uncertainties by
making a simple assumption -- that the relative metal abundances are
roughly solar. In this case, the P {\sc v} $\lambda\lambda$1118,1128
multiplet should have at least 500 times lower optical depth than C
{\sc iv} $\lambda\lambda$1549,1551. Nonetheless, a P {\sc v} BAL is
present in at least half of the well-measured BALQSOs we studied. We
conclude that the strong lines of abundant species like C {\sc iv}
are typically very optically thick. The total BAL column densities
are $N_H\ga 10^{22}$ cm$^{-2}$ (for solar overall metallicity), and
they might be comparable to the X-ray absorbers, of order $10^{23}$
cm$^{-2}$, if the BAL gas is sufficiently ionized. If the column
densities in {\it outflowing} BAL gas are, in fact, as large as the
X-ray absorbers, it would present a serious challenge to models of
radiatively-driven BAL winds.

\end{abstract}

\section{Introduction}

Broad absorption lines (BALs) in quasar spectra identify a dynamic
but still mysterious component quasar environments. These features
clearly form in high velocity winds from the central engines, with
maximum speeds of typically 10,000 km s$^{-1}$ to 30,000 km s$^{-1}$,
but many fundamental questions remain. For example, what are the
total absorbing column densities? Recent X-ray studies have shown
that the columns are typically large, in the range $N_H \approx
10^{23}$ to 10$^{24}$ cm$^{-2}$, but how much of that column density
is actually outflowing with the UV absorption-line gas? Are the flows
radiatively accelerated? What is the geometry, location and ``launch
radius" of the flows, e.g., with respect to the accretion disk and
emission line regions? What are the mass loss rates? What are the
elemental abundances, ionizations, space densities, volume filling
factors, etc. of the absorbing gas?

The most fundamental source of uncertainty (and confusion) in the
analysis of BAL winds is the column densities. For a fully resolved
absorption line that completely covers the emission source(s) along
our line(s) of sight, the measured flux across the line profile,
$F_v$, depends on the optical depth profile, $\tau_v$, by, $F_v =
F_c\, e^{-\tau_v}$, where $F_c$ is the ``continuum" flux (which may
include line emission). This relationship ignores scattered/emitted
flux from the absorbing region itself, but that contribution is
generally considered to be small for BAL winds with small global
covering factors ($\Omega /4\pi \ll 1$, cf. Hamann et al. 1993). The
column density in the measured ion is then, $N_{ion} \propto\
\int\tau_v\, dv$. This analysis applied to the usual UV BALs, such as
Ly$\alpha$, N {\sc v} $\lambda\lambda$1238,1242, Si {\sc iv}
$\lambda\lambda$1393,1404 and C {\sc iv} $\lambda\lambda$1549,1551,
yields total column densities of typically $10^{19}\la
N_H \la
10^{20}$ cm$^{-2}$ --- much lower than those
mentioned above based on X-ray observations.
More surprisingly, this analysis leads to
bizarre elemental abundance ratios, such as, [Si/C] $>$ 0.5 and [C/H]
$\approx 1$ to 2 (where [$x/y$] = $\log(x/y) - \log(x/y)_{\odot}$).
The surprising detections of P {\sc v} $\lambda\lambda$1118,1128 BALs
in sources imply [P/C] $\ga$ 1.6 (see Turnshek 1988, Junkkarinen et
al. 1997, Hamann et al. 1993, Hamann 1998, and reference therein).

How can we understand these strange abundances and reconcile the
vastly different UV and X-ray column densities? One very reasonable
possibility is that the UV and X-ray absorbers are distinct. The
large column density of X-ray absorbing gas might not be outflowing
along with the BAL wind. Perhaps it is more or less stationary in the
quasar rest frame, and the BAL gas is accelerated at larger radii
downstream from the X-ray absorber (e.g., Murray et al. 1995). This
situation could explain vastly different UV and X-ray absorbing
columns, but it cannot explain the strange abundance ratios.

Another possibility is that the analysis of the BALs outlined above
is simply incorrect (or incomplete). In particular, there is growing
evidence that BAL column densities are underestimated by this
analysis because of partial coverage of the background light
source(s). There is unabsorbed flux filling in the bottoms of BAL
troughs, and so the line profiles are {\it not} simply related to the
optical depth by, $F_v = F_c\, e^{-\tau_v}$. Direct evidence for
partial coverage in BALs has come from comparisons of line pairs in
one BALQSO (Arav et al. 19xx). Indirect evidence comes from
flat-bottomed troughs that do not reach zero intensity, and from the
growing numbers of intrinsic ``mini-BALs" and narrow absorption lines
(NALs) whose resolved doublet ratios (e.g., in C {\sc iv}) frequently
require partial coverage (Hamann et al. 1997, Barlow, Hamann \&
Sargent 1997, Arav et al. 1999, Telfer et al. 1998). Changes in the
percent polarization across BAL profiles (e.g., Schmidt \& Hines
1999) might also indicate partial coverage, in this case because of
reflected continuum and/or line emission.

Hamann (1998) argued that the surprisingly strong P {\sc v} BALs are
also evidence for partial coverage. The C {\sc iv} and P {\sc v}
lines form under vary similar physical conditions and their ratio
leads trivially to a lower limit on [P/C]. In the Sun, phosphorus is
$\sim$1000 times less abundant than carbon. With this abundance
ratio, the optical depth in P {\sc v} should be at least $\sim$500
times less than C {\sc iv}. Therefore, if the relative abundances are
even close to solar, e.g. [P/C] $\approx$ 0, the P {\sc v} line
should not be present in BALQSO spectra {\it unless} C {\sc iv} and
other strong lines of abundant elements (such as O {\sc vi}
$\lambda\lambda$1031,1038) are very optically thick. The optical
depths and column densities turn out to be much larger than one would
generally derive from the naive analysis of BALs outlined above.

Below we discuss the implications of large column densities for BAL
winds. We first summarize the findings from a detailed analysis of UV
+ X-ray absorption in a particular BALQSO, PG~1254+047 (Hamann 1998,
Sabra et al. 2001). Then we discuss some preliminary results from an
ongoing study of a larger BALQSO sample (Junkkarinen et al. 2002).

\section{Results for PG 1254+047}

PG~1254+047 ($z_{em} = 1.01$) is particularly interesting because it
has ``detached" BAL troughs; in other words, there is no absorption
near the quasar emission redshift (see Figure 1). The BALs identify a
wind that intersects our line(s) of sight {\it only} at high
blueshifted velocities, from roughly 15,000 to 27,000 km s$^{-1}$.
The strong X-ray absorption in this source (see below) must either be
outflowing at the same high speeds as the BAL gas, or at rest (or at
lower speeds) but producing no significant UV absorption lines.
Either way, we have additional strong constraints on the nature of
the BAL wind.

Hamann (1998) measured a P {\sc v} BAL in PG~1254+047 whose strength
compared to the C {\sc iv} absorption suggests [P/C] $\ga$ 2.2 by the
analysis in \S1 (assuming complete line-of-sight coverage). The data
do not provide direct evidence for partial coverage. However, by
making the reasonable assumption that the relative metal abundances
are {\it roughly} solar (e.g. [P/C] $\approx 0$), Hamann (1998) used
photoionization calculations to show that the line optical depths are
much larger than one would infer simply from the depths of the
observed troughs (\S1). In particular, $\tau_v$(C {\sc iv}) $\ga$ 25
and $\tau_v$(O {\sc vi}) $\ga$ 60. The total column density needed to
produce the measured P {\sc v} is at least $N_H \ga 10^{22}$
cm$^{-2}$ if the overall metallicity is also solar. There is also a
lower limit on the ionization parameter (i.e., the dimensionless
ratio of hydrogen-ionizing photon to hydrogen particle densities at
the illuminated face of the absorbing clouds) of $\log U \ga -0.6$,
based on the absence of low-ionization BALs, such as Mg {\sc ii}
$\lambda\lambda$2796,2804. The total column density could be
substantially larger if the gas is more highly ionized. In
particular, the outflowing gas could have $N_H \approx 10^{23}$
cm$^{-2}$ if $\log U \approx 0.5$. The upper limit on the ionization
and therefore the column density are unknown. Figure 7 in Hamann
(1998) shows the permitted range of $N_H$ values as a function of
$U$.

More recent Chandra X-ray observations by Sabra et al. (2001, see
also Sabra \& Hamann this proceedings) show that, like other BALQSOs,
PG~1254+047 is also severely absorbed in soft X-rays. Sabra et al.
derive a total column density in X-ray absorbing gas of $N_H
\approx 10^{23}$ cm$^{-2}$ if $\log U \approx 0.5$. (It is not
possible to simultaneously derive $N_H$ and $U$ from the existing
X-ray spectrum with only $\sim$47 counts; however, an ionized
absorber does seem to fit the data better than a neutral one.) The
important point here is that the total column density suppressing the
X-rays is consistent with the heavily saturated BALs. How common is
this result?

\begin{figure}
\centerline{\epsfig{figure=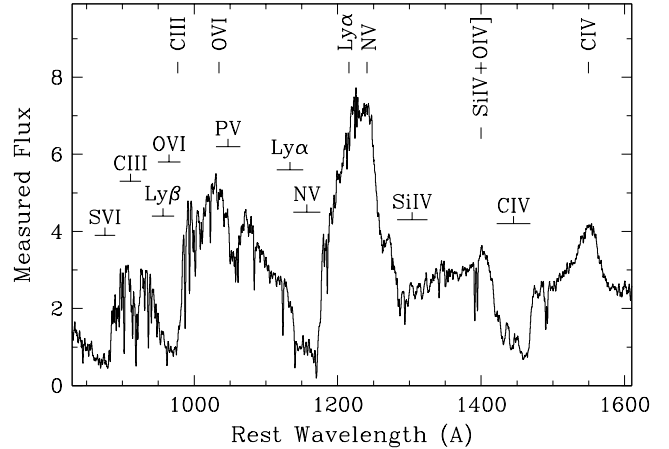,width=5.5cm,angle=0}}
\caption[]{Hubble Space Telescope spectrum of PG~1254+047 showing
its detached BAL troughs. The BALs are labeled just above the
spectrum, while the prominent broad emission lines are marked across
the top. The measured flux has units 10$^{-15}$ ergs/s/cm$^2$/\AA .
See Hamann (1998) for details.}
\end{figure}

\section{A Survey for P {\sc v} in BALQSOs}

To understand the frequency of P {\sc v} absorption in BALQSOs, we
have obtained rest-frame UV spectra for a small sample of 8 BALQSOs
using the Hubble Space Telescope (HST, Junkkarinen et al. 2002). The
sample consists of bright sources with moderate redshifts (so P {\sc
v} $\lambda\lambda$1118,1128 is shifted into the HST wavelength range
while avoiding severe contamination by the Ly$\alpha$ forest that
would occur at higher redshifts). We also tried to select BALQSOs
with relatively narrow line components, so that we can use the
multiplet ratio $\lambda$1118/$\lambda$1128 in P {\sc v} to diagnose
directly the amount of saturation in these lines. To increase the
sample size, we also include spectra from the HST archives of several
other BALQSOs. Importantly, all of the spectra were obtained without
any prejudice regarding the P {\sc v} line strength. Finally, we have
proposed to obtain X-ray spectra for some of the BALQSOs in this
sample to compare directly their UV and X-ray absorbing properties.

The HST spectra are currently under analysis. Detections of the P
{\sc v} BALs can be hampered by blending with nearby lines, such as O
{\sc vi} and C {\sc iii} $\lambda$977. Nonetheless, a preliminary
review of the data indicates that broad P {\sc v} absorption is
present in at least 50\% of the measured sources. With the assumption
that P/C is roughly solar, we again note that the optical depths in P
{\sc v} should be at least 500 times lower than C {\sc iv}. Our
tentative conclusion is that saturation is common in the strong lines
of abundant species, such as C {\sc iv}, N {\sc v} and O {\sc vi},
and therefore, by analogy with PG~1254+047 above, the total column
densities in BAL winds are frequently greater than $10^{22}$
cm$^{-2}$.

\section{What Does this Imply for BAL Wind Dynamics?}

Large column densities in outflowing BAL gas place strong constraints
on wind models that employ radiative acceleration. The equation of
motion for a radial wind emanating from a point source can be written
as,
$$
{{vdv}\over{dr}}\ = \ {{f_L L}\over{4\pi r^2\, c\mu_H m_p\, N_H}}\;
-\; {{GM}\over{r^2}}
$$
where $v$ is the wind velocity at radius $r$, $M$ is the central
black hole mass, $L$ is the quasar luminosity, and $f_L$ is the
fraction of the incident flux that is absorbed or scattered by the
wind (along a given line of sight). Integrating this expression from
an initial ``launch" radius to infinity yields the terminal velocity:
$$
v_{\infty} \ \approx \ 9300\, R_{0.1}^{-1/2}\left({{{f_{0.1}
L_{46}}\over{N_{22}}} \, - \, 0.1 M_8}\right)^{1/2}~~~ {\rm km/s}
$$
where $R_{0.1}$ is the launch radius in units of 0.1 pc, $f_{0.1}$ is
the absorption fraction relative to 10\%, $L_{46}$ is the luminosity
relative to $10^{46}$ ergs/s, $N_{22}$ is the column density relative
to $10^{22}$ cm$^{-2}$, and $M_8$ is the black hole mass relative to
$10^8$ M$_{\odot}$. $L_{46} = 1$ is the Eddington luminosity for a
mass $M_8 \approx 1$, and $f_{0.1} \approx 1$ should be roughly
appropriate for normal high-ionization BALQSOs (Hamann 1998).

BAL winds with total column densities of order $N_H\approx 10^{22}$
cm$^{-2}$ ($N_{22}\approx 1$) should have no trouble reaching a
typical terminal velocity of 15,000 to 20,000 km/s, as long as the
launch radius is $R\la 10^{17}$ cm ($R_{0.1} \leq 0.3$). However, if
the outflows have $N_H\ga 10^{23}$ cm$^{-2}$ ($N_{22}\ga 10$), which
is comparable to the X-ray results and consistent with the measured
BALs in PG~1254+047 (see above), then the radiative force might be
too small to overcome gravity at any launch radius. A gas with
$f_{0.1}\approx 1$ and this total column would remain gravitationally
bound to the black hole. Even if we ignore gravity for a moment, the
launch radius of a flow with $N_H\ga 10^{23}$ cm$^{-2}$ must be very
small, $R\la 10^{16}$ cm, to reach the observed terminal velocities.

This small launch radius would not seem to be a problem, except that
the BAL gas is supposed to reside outside of (at larger radii than)
the broad emission line region (BELR). The main evidence for this
statement comes from observations of N {\sc v} BALs that clearly
absorb the underlying Ly$\alpha$ BEL whenever wind material appears
at the right flow velocity ($\sim$5900 km/s). In a small fraction of
cases, where there is deep absorption setting in abruptly near the
BEL rest velocity, it is again clear that the BAL wind absorbs both
the continuum and broad line radiation. The radius of the Ly$\alpha$
and/or C {\sc iv} BELR, $R_{BELR}$, therefore sets a minimum radius
for the observed BAL wind, $R_{min}$, such that,
$$
R_{min}\ \approx \ R_{BELR} \ \approx \ 0.05\, L_{46}^{0.7} ~ {\rm
pc} \ \approx \ 10^{17} L_{46}^{0.7}  ~ {\rm cm}
$$
where the estimate of $R_{BELR}$ comes from variability studies
(Kaspi et al. (2001). We emphasize that this limit applies to the
{\it observed} BAL wind radius because the launch radius might be
smaller. Certainly in PG~1254+046 (and other BALQSOs with detached
troughs), the wind is launched somewhere out of our line(s) of sight
to the emission sources (requiring a non-radial velocity component --
see the sketches in Ganguly et al. 2001 and Elvis 2000). By the time
the wind intersects our sightline(s), it is already travelling at
speeds $\geq$15,000 km/s. A similar situation explain the sources
with deep BAL absorption near the rest velocity, in this case
requiring a geometry where the highest observed wind speeds occur at
the smallest radii. In any case, it is not really clear what
plausible wind geometry could allow a launch radius that is 10 or
more times smaller than the observed BAL region radius. Moreover, as
we noted above, BAL winds with column densities of order $10^{23}$
cm$^{-2}$ might be gravitationally bound no matter the launch radius.
If the outflowing column densities do prove to be this large, some
other mechanism is likely contributing to the acceleration (e.g.,
Emmering et al. 1992).

\begin{acknowledgements}
We are grateful to the National Science Foundation, the Space
Telescope Science Institute and the Chandra Science Center for grants
supporting this research.
\end{acknowledgements}

\end{document}